\documentclass[aip,12pt,a4paper,preprint,amsmath,amssymb,groupedaddress]{revtex4-2} 

\usepackage{bm}
\usepackage{graphicx}
\graphicspath{figure}

\usepackage{times}
\usepackage{mathptmx}

\DeclareSymbolFont{newfont}{OML}{cmm}{m}{it}
\DeclareMathSymbol{\Varrho}{3}{newfont}{37}

\newcommand{\p}{\partial}

\newcommand{\wt}{\widetilde}

\newcommand{\ave}[1]{{\left<#1\right>}}

\newcommand{\abs}[1]{{\left|#1\right|}}

\newcommand{\distp}{w}

\newcommand{\sol}{{SOL}}

\newcommand{\taud}{\ensuremath{\tau_\text{d}}}

\newcommand{\tauw}{\ensuremath{\tau_\text{w}}}

\newcommand{\aave}{\ensuremath{\ave{a}}}

\newcommand{\vave}{\ensuremath{\ave{v}}}

\newcommand{\ellave}{\ensuremath{\ave{\ell}}}

\newcommand{\tauave}{\ensuremath{\ave{\tau}}}
\newcommand{\Phiave}{\ensuremath{\ave{\Phi}}}

\newcommand{\Phirms}{\ensuremath{\Phi}_\text{rms}}
\newcommand{\Phimin}{\ensuremath{\Phi}_{K_\text{min}}}
\newcommand{\Phimax}{\ensuremath{\Phi}_{K_\text{max}}}

\newcommand{\Kmin}{\ensuremath{K_\text{min}}}
\newcommand{\Kmax}{\ensuremath{K_\text{max}}}

\newcommand{\amin}{\ensuremath{a_\text{min}}}
\newcommand{\amax}{\ensuremath{a_\text{max}}}
\newcommand{\vmin}{\ensuremath{v_\text{min}}}
\newcommand{\vmax}{\ensuremath{v_\text{max}}}
\newcommand{\rmin}{\ensuremath{r_\text{min}}}
\newcommand{\rmax}{\ensuremath{r_\text{max}}}

\newcommand{\ellmin}{\ensuremath{\ell_\text{min}}}
\newcommand{\ellmax}{\ensuremath{\ell_\text{max}}}

\newcommand{\taup}{\ensuremath{\tau_\shortparallel}}

\newcommand{\Eqref}[1]{Eq.~\eqref{#1}}
\newcommand{\Eqsref}[1]{Eqs.~\eqref{#1}}
\newcommand{\Figref}[1]{Fig.~\ref{#1}}

\newcommand{\Secref}[1]{Sec.~\ref{#1}}

\newcommand{\Appref}[1]{App.~\ref{#1}}

\begin{document}

\title{Stochastic modelling of blob-like plasma filaments in the scrape-off layer: Theoretical foundation}

\author{J.~M.~Losada}
\email{juan.m.losada@uit.no}

\author{A.~Theodorsen}
\email{audun.theodorsen@uit.no}

\author{O.~E.~Garcia}
\email{odd.erik.garcia@uit.no}

\affiliation{Department of Physics and Technology, UiT The Arctic University of Norway, N-9037 Troms{\o}, Norway}

\date{\today}

\begin{abstract}
A stochastic model for a super-position of uncorrelated pulses with a random distribution of amplitudes, sizes and velocities is presented. The pulses are assumed to move radially with fixed shape and amplitudes decaying exponentially in time due to linear damping. The pulse velocities are taken to be time-independent but randomly distributed. The implications of a distribution of pulse amplitudes, sizes and velocities are investigated. Closed-form expressions for the cumulants and probability density functions for the process are derived in the case of exponential pulses and a discrete uniform distribution of pulse velocities. The results describe many features of the boundary region of magnetically confined plasmas such as high average particle densities, broad and flat radial profiles, and intermittent large-amplitude fluctuations. The stochastic model elucidates how these phenomena are related to the statistical properties of blob-like structures. In particular, the presence of fast pulses generally leads to flattened far scrape-off layer profiles and enhanced intermittency, which amplifies plasma--wall interactions.
\end{abstract}

\maketitle

\section{Introduction}

Magnetically confined fusion plasmas in toroidal geometry rely on a poloidal divertor topology in order to control plasma exhaust.\cite{stangeby_plasma_2000,fundamenski_power_2014,krasheninnikov_edge_2020,militello_bonundary_2023} Plasma entering the scrape-off layer (\sol) from the core will flow along magnetic field lines to the remote divertor chamber, which is specifically designed to handle the exhaust of particles and heat. This is supposed to avoid strong plasma--wall contact in the main chamber, which is located close to the core plasma. However, experiments have shown that cross-field plasma transport is generally significant and may even be dominant, leading to detrimental plasma interactions with the main chamber walls.\cite{stangeby_plasma_2000,fundamenski_power_2014,krasheninnikov_edge_2020,militello_bonundary_2023,pitts_material_2005,lipschultz_plasma-surface_2007,brooks_plasma-surface_2009,marandet_transport_2011,birkenmeier_filament_2015,meyer_overview_2017}

Measurements on numerous tokamak devices have demonstrated that as the core plasma density increases, the particle density in the \sol\ becomes higher and plasma--wall interactions increase.\cite{asakura_sol_1997,labombard_experimental_1997,labombard_cross-field_2000,labombard_particle_2001,lipschultz_investigation_2002,lipschultz_comparison_2005,whyte_magnitude_2005,carralero_implications_2015,militello_characterisation_2016,wynn_investigation_2018,antar_experimental_2001,boedo_transport_2001,rudakov_fluctuation_2002,antar_universality_2003,boedo_intermittent_2003,rudakov_far_2005-1,kirnev_comparison_2005,labombard_evidence_2005,horacek_overview_2005,garcia_interchange_2006,garcia_fluctuations_2007,garcia_collisionality_2007,antar_turbulence_2008,tanaka_statistical_2009,silva_intermittent_2009,horacek_interpretation_2010,yan_statistical_2013,carralero_experimental_2014,carralero_study_2017,vianello_modification_2017,carralero_recent_2017,kube_statistical_2019,vianello_scrape-off_2020,stagni_dependence_2022,theodorsen_scrape-off_2016,garcia_sol_2017,walkden_interpretation_2017,kuang_plasma_2019,kube_comparison_2020,zurita_stochastic_2022} The particle density profile in the \sol\ typically exhibits a two-layer structure, commonly referred to as a density shoulder. Close to the magnetic separatrix, in the so-called near \sol, it has a steep exponential decay and moderate fluctuation levels. Beyond this region, in the so-called far \sol, the profile has an exponential decay with a much longer scale length. As the core plasma density increases, the profile scale length in the far \sol\ becomes longer, referred to as profile flattening, and the break point between the near and the far \sol\ moves radially inwards, referred to as profile broadening. When the empirical discharge density limit is approached, the far \sol\ profile effectively extends all the way to the magnetic separatrix or even inside it.

The boundary region of magnetically confined plasmas is generally in an inherently fluctuating state. Single point measurements of the particle density in the far \sol\ reveal frequent occurrence of large-amplitude bursts and relative fluctuation levels of order unity.\cite{antar_experimental_2001,boedo_transport_2001,rudakov_fluctuation_2002,antar_universality_2003,boedo_intermittent_2003,rudakov_far_2005-1,kirnev_comparison_2005,labombard_evidence_2005,horacek_overview_2005,garcia_interchange_2006,garcia_fluctuations_2007,garcia_collisionality_2007,antar_turbulence_2008,tanaka_statistical_2009,silva_intermittent_2009,horacek_interpretation_2010,yan_statistical_2013,carralero_experimental_2014,carralero_study_2017,vianello_modification_2017,carralero_recent_2017,kube_statistical_2019,vianello_scrape-off_2020,stagni_dependence_2022,theodorsen_scrape-off_2016,garcia_sol_2017,walkden_interpretation_2017,kuang_plasma_2019,kube_comparison_2020,zurita_stochastic_2022,garcia_burst_2013,garcia_intermittent_2013,garcia_intermittent_2015,kube_fluctuation_2016,theodorsen_relationship_2017,kube_intermittent_2018,garcia_intermittent_2018,theodorsen_universality_2018,bencze_characterization_2019,zweben_temporal_2022} The large-amplitude fluctuations, identified in the \sol\ of all tokamaks and in all confinement regimes, are attributed to radial motion of coherent structures through the \sol\ and towards the main chamber wall. These structures are observed as magnetic-field-aligned filaments of excess particles and heat as compared to the ambient plasma, commonly referred to as blobs.\cite{zweben_search_1985,zweben_edge_2002,terry_observations_2003,zweben_high-speed_2004,terry_velocity_2005,grulke_radially_2006,terry_spatial_2009,ben_ayed_inter-elm_2009,maqueda_intermittency_2011,agostini_edge_2011,banerjee_statistical_2012,kube_blob_2013,grulke_experimental_2014,fuchert_blob_2014,zweben_blob_2016} This leads to broad and flat far \sol\ profiles and enhanced levels of plasma interactions with the main chamber walls that may be an issue for the next generation magnetic confinement experiments.

At the outboard mid-plane region, localized blob-like structures get charge polarized due to vertical magnetic gradient and curvature drifts. The resulting electric field leads to radial motion of the filament structures towards the main chamber wall. The strongly non-linear advection results in an asymmetric shape with a steep front and a trailing wake. Single-point measurements record the filaments as an asymmetric, two-sided exponential pulse function.\cite{theodorsen_scrape-off_2016,garcia_sol_2017,walkden_interpretation_2017,kuang_plasma_2019,kube_comparison_2020,zurita_stochastic_2022,garcia_burst_2013,garcia_intermittent_2013,garcia_intermittent_2015,kube_fluctuation_2016,theodorsen_relationship_2017,kube_intermittent_2018,garcia_intermittent_2018,theodorsen_universality_2018,bencze_characterization_2019,zweben_temporal_2022}
The radial filament velocity depends on the blob size and amplitude as well as plasma parameters and takes a wide range of values. This dependence has been extensively explored theoretically and by numerical simulations of isolated filament structures in various plasma parameter regimes.\cite{krasheninnikov_scrape_2001,dippolito_cross-field_2002,bian_blobs_2003,garcia_mechanism_2005,garcia_radial_2006,myra_collisionality_2006,dippolito_convective_2011,madsen_influence_2011,kube_velocity_2011,higgins_determining_2012,manz_filament_2013,halpern_three-dimensional_2014,easy_three_2014,easy_investigation_2016,held_influence_2016,pecseli_solvable_2016,kube_amplitude_2016,walkden_dynamics_2016,wiesenberger_unified_2017,ross_nature_2019} Filament velocity scaling properties have also been investigated experimentally, showing correlations with other blob parameters and plasma parameters.\cite{zweben_search_1985,zweben_edge_2002,terry_observations_2003,zweben_high-speed_2004,terry_velocity_2005,grulke_radially_2006,terry_spatial_2009,ben_ayed_inter-elm_2009,maqueda_intermittency_2011,agostini_edge_2011,banerjee_statistical_2012,kube_blob_2013,grulke_experimental_2014,fuchert_blob_2014,zweben_blob_2016} It follows that stochastic modelling of the intermittent fluctuations in the \sol\ must necessarily include a distribution of filament velocities.

In previous experimental investigations based on second-long, single-point measurement data time series, the fundamental statistical properties of the plasma fluctuations in the far \sol\ have been identified.\cite{theodorsen_scrape-off_2016,garcia_sol_2017,walkden_interpretation_2017,kuang_plasma_2019,kube_comparison_2020,zurita_stochastic_2022,garcia_burst_2013,garcia_intermittent_2013,garcia_intermittent_2015,kube_fluctuation_2016,theodorsen_relationship_2017,kube_intermittent_2018,garcia_intermittent_2018,theodorsen_universality_2018,bencze_characterization_2019,zweben_temporal_2022} It has been demonstrated that the fluctuations can be described as a super-position of uncorrelated, exponential pulses with an exponential distribution of pulse amplitudes, referred to as a filtered Poisson process.\cite{garcia_stochastic_2012,kube_convergence_2015,theodorsen_level_2016,theodorsen_level_2018,theodorsen_probability_2018,theodorsen_statistical_2017,garcia_auto-correlation_2017,garcia_stochastic_2016} For such a stochastic process the probability density function is a Gamma distribution with the scale parameter given by the average pulse amplitude and the shape parameter given by the ratio of the average pulse duration and waiting times.\cite{garcia_stochastic_2012,kube_convergence_2015,theodorsen_level_2016,theodorsen_level_2018,theodorsen_probability_2018,theodorsen_statistical_2017,garcia_auto-correlation_2017,garcia_stochastic_2016} Moreover, it follows that the auto-correlation function has an exponential tail and the frequency power spectral density has a Lorentzian shape.\cite{garcia_stochastic_2016,garcia_auto-correlation_2017,theodorsen_statistical_2017} Both the underlying assumptions of the model and its predictions are found to be in excellent agreement with experimental measurements.\cite{theodorsen_scrape-off_2016,garcia_sol_2017,walkden_interpretation_2017,kuang_plasma_2019,kube_comparison_2020,zurita_stochastic_2022,garcia_burst_2013,garcia_intermittent_2013,garcia_intermittent_2015,kube_fluctuation_2016,theodorsen_relationship_2017,kube_intermittent_2018,garcia_intermittent_2018,theodorsen_universality_2018,bencze_characterization_2019,zweben_temporal_2022}

Recently, the statistical description of single-point measurements was extended to describe the radial variation of the average \sol\ profile due to the motion of blob-like filament structures with a random distribution of sizes and velocities.\cite{garcia_stochastic_2016,militello_scrape_2016,militello_relation_2016, paikina_2022} This reveals how the average profile and its radial variation depend on the filament statistics. In particular, if all filaments have the same velocity, the radial e-folding length is given by the product of the radial filament velocity and the parallel transit time to the divertor targets. In this presentation, we extended and complement this statistical analysis by a systematic study of randomly distributed filament amplitudes, sizes and velocities, and correlations between these quantities. The filaments are assumed to move radially outwards with fixed shape and amplitudes decaying exponentially in time due to linear damping. The velocities are taken to be time-independent but may be correlated with other filament parameters. The combination of linear damping and a random distribution of velocities is shown to significantly modify the average profiles as well as the fluctuations in the process. The results presented here extend previous work by including predictions for higher-order moments, in particular skewness and flatness profiles. Closed-form analytical expressions are obtained in the case of a discrete uniform distribution of pulse velocities.

This paper is the first in a sequence, presenting extensions of the filtered Poisson process to describe the radial motion of pulses including linear damping due to parallel drainage in the scrape-off layer. This first paper defines the theoretical framework and gives a derivation of all the general results for the case of time-independent pulse velocities and provides closed-form expressions for the relevant statistical averages in the case of a discrete uniform distribution of pulse velocities. Follow-up papers will address various continuous distributions of pulse velocities, cases where the pulse velocity depends on the pulse amplitude, time-dependent pulse velocities, the correlation functions and frequency and wave number spectra of the fluctuations, and extensions to several spatial dimensions.

The organization of this paper is as follows. In \Secref{sec.model} we present the stochastic model describing a super-position of pulses with a random distribution of and correlations between amplitudes, sizes and velocities. In \Secref{sec.timeindependent} we derive general expressions for the cumulants and discuss how the combination of radial motion and linear damping influences the statistical properties of the fluctuations. In \Secref{sec.numerical} we present closed-form expressions for the radial profile of the lowest order statistical moments and probability distributions for the case of a discrete uniform distribution of pulse velocities. A discussion of the results in the context of blob-like filament structures at the boundary of magnetically confined plasmas is presented in \Secref{sec.discussion} and the conclusions and an outlook are given in \Secref{sec.conclusions}. The paper is complemented by four appendices. In \Appref{app.two.exp} general results are presented for the case of two-sided exponential pulses. End effects in realizations of the process are considered in \Appref{app.fpp}. Appendix~\ref{app.cumulants} discusses limitations on the existence of cumulants. Implications of a discrete uniform distribution of pulse sizes are presented in \Appref{app.size}. Finally, the generalization to a non-uniform discrete velocity distribution is given in \Appref{app.nonunif}.

\section{Stochastic model}\label{sec.model}

In this section, the stochastic process is presented, describing a super-position of uncorrelated pulses which do not interact with each other. It is demonstrated that this is a generalization of a filtered Poisson process with particularly transparent results obtained for an exponential pulse function.

\subsection{Super-position of pulses}

Consider the evolution of an individual pulse $\phi(x,t)$, which is assumed to follow an advection equation on the form
\begin{equation}\label{eq.dphikdt}
    \frac{\p\phi}{\p t} + v\,\frac{\p \phi}{\p x} + \frac{\phi}{{\taup}} = 0 ,
\end{equation}
where $v$ is the pulse velocity along the radial axis $x$. The last term on the left-hand side describes linear damping with e-folding time $\taup$, which is assumed to be constant in time and independent of the pulse parameters. The linear damping originates from the parallel drainage of plasma along the magnetic field lines, hence the subscript $\shortparallel$. The pulse velocity $v$ will in the following be assumed to be positive and time-independent. As initial condition, we take that the pulse $\phi$ is assumed to arrive at the reference position $x=0$ at time $t=0$,
\begin{equation}
    \phi(x,0) = a \varphi\left( \frac{x}{\ell} \right) ,
\end{equation}
where $a$ and $\ell$  are the pulse amplitude and size, respectively. The non-dimensional pulse function $\varphi(\theta)$ is taken to be the same for all events and satisfies the normalization constraint
\begin{equation}
    \int_{-\infty}^{\infty} \text{d}\theta\,\abs{\varphi(\theta)} = 1 .
\end{equation}
For later reference, we define the integral of the $n$'th power of the pulse function as
\begin{equation}\label{eq.In}
    I_n = \int_{-\infty}^{\infty} \text{d}\theta\,[\varphi(\theta)]^n .
\end{equation}
For a non-negative pulse function, it follows that $I_1=1$. Applying the method of characteristics to the differential equation \eqref{eq.dphikdt} leads to the general solution
\begin{equation}\label{eq.phik}
    \phi(x,t) = A(t) \varphi\left( \frac{x-v t}{\ell} \right) ,
\end{equation}
where the pulse amplitude evolution is determined by
\begin{equation}\label{eq.amplitude}
    A(t) = a \exp \left( - \frac{t}{\taup} \right) .
\end{equation}
Equation~\eqref{eq.phik} determines the pulse evolution for given amplitude $a$, size $\ell$ and velocity $v$. The pulse moves radially without change in shape but with an amplitude that decays exponentially in time due to the linear damping.

Consider now the stochastic process $\Phi_K(x,t)$ given by a super-position of $K$ uncorrelated and spatially localized pulses,
\begin{align} \label{eq.PhiK}
    \Phi_K(x,t) & = \sum_{k=1}^{K(T)} \phi_k(x,t-s_k) \notag
    \\
    & = \sum_{k=1}^{K(T)} a_k \exp\left( - \frac{t-s_k}{\taup} \right) \varphi\left( \frac{x-v_k(t-s_k)}{\ell_k} \right) ,
\end{align}
where each pulse $\phi(x,t)$ satisfies \Eqref{eq.dphikdt}. In the following, the subscript $k$ on the random variables $a$, $v$ and $\ell$ will be suppressed when possible for simplicity of notation. Each pulse is located at $x=0$ at the arrival time $s$. All other pulse parameters are assumed to be independent of the arrival times. The arrival times $s$ are furthermore assumed to be independent and uniformly distributed on an interval of duration $T$, that is, their probability distribution function is
\begin{equation}\label{eq.p_tk.uniform}
    P_s(s_k) =
        \begin{cases}
            1/T , & \abs{t} \leq T/2 , \\
            0 , & \abs{t} > T/2 .
        \end{cases}
\end{equation}
With these assumptions, the probability that there are exactly $K$ pulse arrivals at $x=0$ during any interval of duration $T$ is given by the Poisson distribution
\begin{equation} \label{eq.poisson}
    P_K(K;T) = \frac{1}{K!}\left(\frac{T}{\tauw}\right)^K\exp\left(-\frac{T}{\tauw} \right) ,
\end{equation}
where $\tauw$ is the average pulse waiting time at the reference position $x=0$. The average number of pulses in realizations of duration $T$ is
\begin{equation}
    \ave{K} = \sum_{K=0}^\infty K P_K(K;T) = \frac{T}{\tauw} ,
\end{equation}
where, here and in the following, angular brackets denote the ensemble average of a random variable over all its arguments. From the Poisson distribution, it follows that the waiting time between two subsequent pulses is exponentially distributed. It is emphasized that all pulse parameters and their correlations are specified at the reference position $x=0$. In particular, the Poisson property of the process is defined for this reference position but does not necessarily hold for other radial positions. This will be discussed further in \Secref{sec.fpp} and \Appref{app.fpp}.

\subsection{Exponential pulses}

The exponential amplitude modulation due to linear damping in \Eqref{eq.amplitude} suggests that particularly simple expressions may be obtained for a similar dependence in the pulse function. We thus consider the case of a one-sided exponential pulse function,
\begin{equation}\label{eq:pulse-one-sided}
    \varphi(\theta) =
        \begin{cases}
            \exp{(\theta)}, &  \theta \leq 0 , \\
            0, &  \theta > 0 .
        \end{cases}
\end{equation}
In the following sections, this one-sided exponential pulse function will be used to demonstrate the fundamental properties of the process and in order to calculate closed-form expressions for moments and distribution functions. It should be noted that for exponential pulses, the integral $I_n=1/n$. The generalization of the following results to two-sided exponential pulses is discussed in \Appref{app.two.exp}.

\begin{figure}[t]
\centering
\includegraphics[width=8cm]{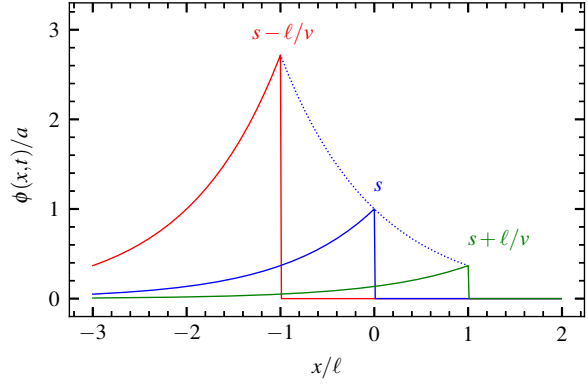}
\caption{Radial variation of a one-sided exponential pulse at the arrival time $s$ and one radial transit time $\ell/v$ before and after the arrival at $x=0$. The dotted line shows the radial variation of the pulse amplitude due to linear damping.}
\label{fig.onesided1}
\end{figure}

The relevant parameters of the process given by \Eqref{eq.PhiK} for a one-sided exponential pulse function are presented in \Figref{fig.onesided1}. Here the radial variation of the pulse is shown for the time of arrival $s$ at $x=0$ as well as one radial transit time $\ell/v$ before and after this arrival time. Due to linear damping, the amplitude decreases exponentially in time as the pulse moves along the radial axis, indicated by the dotted line in the figure.

At the reference position, $x=0$, the process is given by
\begin{equation}\label{eq.PhiK.solution.x0}
    \Phi_K(0,t) = \sum_{k=1}^{K(T)} a_k\exp{\left( - \frac{t-s_k}{\taup} \right)} \varphi\left( - \frac{v_k(t-s_k)}{\ell_k} \right) .
\end{equation}
For the exponential pulse function defined by \Eqref{eq:pulse-one-sided} it is straightforward to show that this process can be written as
\begin{equation}\label{eq.PhiK.solution.lambda}
    \Phi_K(0,t) = \sum_{k=1}^{K(T)} a_k \varphi\left( - \frac{t-s_k}{\tau_k} \right) ,
\end{equation}
where the pulse duration is the harmonic mean of the linear damping time and the radial transit time $\ell/v$,
\begin{equation}\label{eq.tauk.onesided}
    \tau = \frac{\taup\ell}{v\taup+\ell} .
\end{equation}
The average pulse duration is denoted by $\taud=\tauave$ and is clearly influenced by a distribution of pulse sizes and velocities. In the absence of linear damping, the pulse duration is just the radial transit time, $\ell/v$. Further discussions of the pulse duration are given in \Secref{sec.ave.ti}.

\subsection{Filtered Poisson process}\label{sec.fpp.x0}

The process at the reference position $x=0$ describes a super-position of uncorrelated, exponential pulses given by \Eqref{eq.PhiK.solution.lambda}. When all pulses have the same duration $\taud$, the process can be written as a convolution or filtering of the pulse function with a train of delta pulses,\cite{theodorsen_universality_2018}
\begin{align}
    \Phi_K(0,t) & = \int_{-\infty}^{\infty} \text{d}\theta\,\varphi\left( \frac{t}{\taud}-\theta \right) \mathcal{F}_K(\theta) \notag
    \\
    & = (\varphi * \mathcal{F}_K)\left( \frac{t}{\taud} \right) ,
\end{align}
where the forcing is
\begin{equation}
    \mathcal{F}_K(\theta) = \sum_{k=1}^{K(T)} a_k \delta\left( \theta-\frac{s_k}{\taud} \right) .
\end{equation}
This is therefore commonly referred to as a filtered Poisson process. More generally, for the process given by \Eqref{eq.PhiK.solution.lambda} with a random distribution of all pulse parameters, the ratio of the average pulse duration and waiting times,
\begin{equation}
    \gamma = \frac{\taud}{\tauw} ,
\end{equation}
determines the degree of pulse overlap and is referred to as the intermittency parameter of the process.\cite{garcia_stochastic_2012}

In the case of an exponential pulse function and exponentially distributed pulse amplitudes with mean value $\aave$, which for positive $a$ is given by
\begin{equation}\label{eq.pa.exp}
    \aave P_a(a) = \exp{\left( - \frac{a}{\aave} \right)} ,
\end{equation}
the raw amplitude moments are $\langle{a^n}\rangle=n!\aave^n$ and the stationary probability density function for $\Phi_K(0,t)$ is given by a Gamma distribution with shape parameter $\gamma$ and scale parameter $\aave$. For positive $\Phi$ this distribution can be written as\cite{garcia_stochastic_2012,kube_convergence_2015,theodorsen_level_2016,theodorsen_level_2018,theodorsen_probability_2018,theodorsen_statistical_2017,garcia_auto-correlation_2017,garcia_stochastic_2016}
\begin{equation}\label{eq.PPhi.Gamma}
    \aave P_\Phi(\Phi) = \frac{1}{\Gamma(\gamma)}\left( - \frac{\Phi}{\aave} \right)^{\gamma-1}\exp{\left( - \frac{\Phi}{\aave} \right)} ,
\end{equation}
with mean value $\Phiave=\gamma\aave$ and variance $\Phirms^2=\gamma\aave^2$. The intermittency parameter $\gamma$ determines the shape of the distribution, resulting in a high relative fluctuation level as well as skewness and flatness moments in the case of weak pulse overlap for small $\gamma$. The Gamma probability density function holds for any distribution of pulse durations but assumes that the pulse amplitudes and durations are independent.\cite{garcia_auto-correlation_2017,theodorsen_probability_2018}

\section{Moments of the process}\label{sec.timeindependent}

In this section, we present derivations of the mean value, the characteristic function, cumulants and the lowest order statistical moments for a sum of uncorrelated pulses given by \Eqref{eq.PhiK}. Particular attention is devoted to mechanisms for radial variation of moments and intermittency of the process.

\subsection{Average radial profile}\label{sec.ave.ti}

Let us first consider the average of the process $\Phi_K(x,t)$. The arrival times $s$ are taken to be independent of the other pulse parameters. Thus, we first perform the average of each pulse over the arrival times,
\begin{equation}\label{eq.avephi1.ti}
    \ave{\phi(x,t-s)} = \frac{1}{T}\ave{\int_{-T/2}^{T/2} \text{d}s\,a \exp{\left( - \frac{t-s}{\taup} \right)} \varphi\left( \frac{x-v(t-s)}{\ell} \right)} ,
\end{equation}
where the angular brackets denote an average over all amplitudes, sizes and velocities with the $k$ subscript suppressed for simplicity of notation. Neglecting end effects by taking the integration limits for $s$ to infinity and changing the integration variable to $\theta=[x-v(t-s)]/\ell$ gives
\begin{equation}
    \ave{\phi}(x) = \frac{1}{T}\ave{\frac{a \ell}{v}\exp{\left( -\frac{x}{v\taup}\right)} \int_{-\infty}^{\infty} \text{d}\theta \exp{\left(\frac{\theta \ell}{v\taup}\right)} \varphi(\theta)} .
\end{equation}
Given that the pulses are uncorrelated, the average of the conditional process with exactly $K$ pulses is given by $\langle{\Phi_K}\rangle=K\langle{\phi(x,t-s)}\rangle$. Therefore, averaging over the number of pulses gives the general result
\begin{align}\label{eq.mean.general}
    \ave{\Phi}(x) & = \sum_{K=0}^\infty \ave{\Phi_K} P_K(K;T)
    \\ \notag
    & = \frac{1}{\tauw}\ave{\frac{a \ell}{v}\exp{\left( -\frac{x}{v\taup}\right)} \int_{-\infty}^{\infty} \text{d}\theta \exp{\left(\frac{\theta \ell}{v\taup}\right)} \varphi(\theta)} .
\end{align}
In the absence of linear damping, the mean value does not depend on the radial coordinate and is given by $\Phiave=\langle{a\ell I_1/v}\rangle/\tauw$ for any joint distribution between pulse amplitudes, sizes and velocities.

In the case where all pulses have the same velocity, it follows that the average radial profile is exponential with a length scale given by the product of the radial velocity and the linear damping time,
\begin{equation} \label{eq.expprof}
    \ave{\Phi}(x) = \frac{1}{\tauw}\ave{\frac{a \ell}{v} \int_{-\infty}^{\infty} \text{d}\theta \exp{\left(\frac{\theta \ell}{v\taup}\right)} \varphi(\theta)} \exp{\left( -\frac{x}{v\taup}\right)} .
\end{equation}
The exponential profile obviously follows from the combination of radial motion and linear damping of the pulses. More generally, it is clear from \Eqref{eq.mean.general} that a random distribution of pulse velocities will make the average radial profile non-exponential. This will be further investigated in \Secref{sec.numerical}.

\begin{figure}[t]
\centering
\includegraphics[width=8cm]{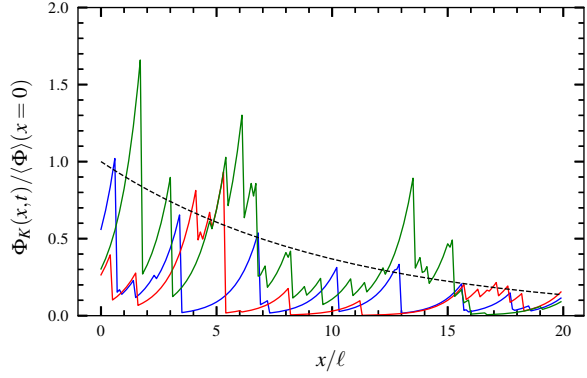}
\caption{Super-position of one-sided exponential pulses with an exponential amplitude distribution and a degenerate distribution of pulse sizes and velocities. The linear damping is given by $v\taup/\ell=10$ and the intermittency parameter by $v\tauw/\ell=1$. Different colours represent various realizations of the process and the dashed line is the predicted exponential radial profile given by \Eqref{eq.Phi.ti.case}.}
\label{fig.realization}
\end{figure}

For the exponential pulse function defined by \Eqref{eq:pulse-one-sided} and any distribution of amplitudes, sizes and velocities, we obtain the average profile
\begin{equation}\label{eq.Phi.ti}
    \ave{\Phi}(x) = \frac{1}{\tauw}\ave{a\tau\exp{\left( -\frac{x}{v\taup}\right)}} ,
\end{equation}
where the pulse duration $\tau$ is given by \Eqref{eq.tauk.onesided}. In the case of a degenerate distribution of the pulse velocities the average radial profile is exponential,\cite{garcia_stochastic_2016,militello_scrape_2016,militello_relation_2016}
\begin{equation}\label{eq.Phi.ti.case}
    \ave{\Phi}(x) = \frac{\ave{a\tau}}{\tauw}\,\exp{\left( - \frac{x}{v\taup} \right)} .
\end{equation}
If additionally, the pulse sizes are uncorrelated with the amplitudes, the prefactor is given by $\aave\taud/\tauw$ with $\taud$ the average pulse duration. Realizations of this process with an exponential amplitude distribution and fixed pulse sizes and velocities are presented in \Figref{fig.realization}. The pulses cause large-amplitude fluctuations to the average radial profile which will now be quantified with cumulants and higher-order moments.

\subsection{Cumulants and moments}\label{sec.cumulants}

The characteristic function for the random variable $\Phi_K$ at the radial position $x$ is the Fourier transform of the probability density function and is given by $C_{\Phi_K}(u;x)=\langle{\exp{(iu\Phi_K)}}\rangle$. The characteristic function for a sum of independent random variables is the product of their individual characteristic functions. Since all pulses $\phi(x,t-s)$ are by assumption independent, and each of the parameters $a$, $\ell$ and $v$ are identically distributed, the characteristic function for the process is given by
\begin{equation}
    C_{\Phi_K}(u;x) = \prod_{k=1}^K C_\phi(u;x) = [C_\phi(u;x)]^K ,
\end{equation}
where we have defined the characteristic function for an individual pulse as
\begin{equation}
    C_\phi(u;x) = \ave{\exp(iu\phi)} ,
\end{equation}
with $\phi_k(x,t-s)$ given by \Eqref{eq.phik} and the average is to be taken over arrival times $s$ and the other randomly distributed pulse parameters.

The conditional probability distribution function of $\Phi_K$ for fixed $K$ is
\begin{equation}
    P_{\Phi_K}(\Phi_K|K) = \frac{1}{2\pi} \int_{-\infty}^{\infty} \text{d}u\,\exp{(iu\Phi_K)} [C_\phi(u,x)]^K .
\end{equation}
Using that $K$ is Poisson distributed as defined by \Eqref{eq.poisson} we have
\begin{align}
    P_\Phi(\Phi) & = \sum_{K=0}^{\infty} P_K(K;T)P_{\Phi_K}(\Phi_K|K)
    \\ \notag
    & = \frac{1}{2\pi} \int_{-\infty}^{\infty} \text{d}u\,\exp{(iu\Phi)} \exp{\left( \frac{T}{\tauw}[C_\phi(u;x)-1]\right)} .
\end{align}
The expression inside the last exponential function can be identified as the logarithm of the characteristic function $C_\Phi(u;x)$, given by
\begin{equation}
    \ln{C_\Phi} = \frac{T}{\tauw}(C_\phi-1) = \frac{1}{\tauw} \ave{ \int_{-T/2}^{T/2} \text{d}s\,\left[ \exp{(iu\phi)} - 1 \right] } ,
\end{equation}
where the averaging in the last expression is over all pulse amplitudes, sizes and velocities. Neglecting end effects by extending the integration limits over pulse arrivals $s$ to infinity and expanding the exponential function we can write
\begin{equation} \label{eq:lnCphi}
    \ln{C_\Phi} = \frac{1}{\tauw} \ave{ \int_{-\infty}^{\infty} \text{d}s\,\left[ \sum_{n=1}^\infty\frac{(iu\phi)^n}{n!} \right] } .
\end{equation}
The statistical moments are directly related to the cumulants $\kappa_n$, which are defined as the coefficients in the expansion of the logarithm of the characteristic function,
\begin{equation}
    \ln{C_\Phi} = \sum_{n=1}^\infty \frac{\kappa_n(iu)^n}{n!} .
\end{equation}
A comparison with \Eqref{eq:lnCphi} gives the cumulants,
\begin{equation}
    \kappa_n(x) = \frac{1}{\tauw} \ave{ \int_{-\infty}^{\infty} \text{d}s [\phi(x,t-s)]^n } .
\end{equation}
Following a similar procedure as for calculating the average radial profile we obtain the general expression for the cumulants,
\begin{equation}\label{eq.cumulants.general}
    \kappa_n(x) = \frac{1}{\tauw}\ave{\frac{a^n\ell}{v}\exp{\left( -\frac{nx}{v\taup}\right)} \int_{-\infty}^{\infty} \text{d}\theta\,\exp{\left(\frac{n\theta \ell}{v\taup}\right)} \left[ \varphi(\theta) \right]^n} .
\end{equation}
In the case of a degenerate distribution of pulse velocities, the cumulants decrease exponentially with radius. The exponential profile is clearly modified by a distribution of pulse velocities.

From the cumulants, the lowest order moments of $\Phi$ are readily obtained. A formal power series expansion shows that the characteristic function $C_{\Phi}(u;x)=\langle{\exp(iu\Phi)}\rangle$ is related to the raw moments $\ave{\Phi^n}$,
\begin{equation}
    C_{\Phi}(u;x) = 1 + \sum_{n=1}^\infty \ave{\Phi}^n\,\frac{(iu)^n}{n!} .
\end{equation}
The first order cumulant is just the mean value, $\kappa_1=\Phiave$, while the second order cumulant is the variance of the process, $\kappa_2=\Phirms^2=\langle{(\Phi-\Phiave)^2}\rangle$, with
\begin{equation}
    \Phirms^2 = \frac{1}{\tauw}\ave{ \frac{a^2\ell}{v}\exp{\left( -\frac{2x}{v\taup}\right)} \int_{-\infty}^{\infty} \text{d}\theta\,\exp{\left(\frac{2\theta \ell}{v\taup}\right)} \left[ \varphi(\theta) \right]^2 } .
\end{equation}
The lowest order centered moments $\mu_n=\langle{(\Phi-\Phiave)^n}\rangle$ are related to the cumulants by the relations $\mu_2=\kappa_2$, $\mu_3=\kappa_3$ and $\mu_4=\kappa_4+3\kappa_2^2$. The skewness and flatness moments are defined by
\begin{align} \label{eq.highorder}
    S_\Phi & = \frac{\ave{(\Phi-\Phiave)^3}}{\Phirms^3} = \frac{\kappa_3}{\kappa_2^{3/2}} ,
    \\
    F_\Phi & = \frac{\ave{(\Phi-\Phiave)^4}}{\Phirms^4} - 3 = \frac{\kappa_4}{\kappa_2^2} .
\end{align}
In the following, we will occasionally use a scaled variable defined by
\begin{equation}\label{Phiscaled}
    \wt{\Phi}(x,t) = \frac{\Phi(x,t)-\Phiave(x)}{\Phirms(x)} ,
\end{equation}
normalized such as to have zero mean and unit standard deviation at all radial positions.

In the absence of linear damping, the cumulants and moments do not depend on the radial coordinate and are given by $\kappa_n=\langle{a^n\ell I_n/v}\rangle/\tauw$. For an exponential pulse function, for which $I_n=1/n$, with exponentially distributed amplitudes that are independent of the pulse duration $\tau=\ell/v$, for which $\langle{a^n}\rangle=n!\aave$, the cumulants simplify to $\kappa_n=(n-1)!\gamma\aave^n$, where $\gamma=\taud/\tauw$ and $\taud$ is the average pulse duration. This is nothing but the cumulants of a Gamma distribution with scale parameter $\aave$ and shape parameter $\gamma$, which is the case discussed in \Secref{sec.fpp.x0}.

In the case of a degenerate distribution of pulse velocities, the variance is given by
\begin{equation}
    \Phirms^2 = \frac{1}{\tauw}\ave{ \frac{a^2\ell}{v} \int_{-\infty}^{\infty} \text{d}\theta\,\exp{\left(\frac{2\theta \ell}{v\taup}\right)} \left[ \varphi(\theta) \right]^2 } \exp{\left( -\frac{2x}{v\taup}\right)} .
\end{equation}
It follows that the relative fluctuation level as well as the skewness and flatness moments do not depend on the radial coordinate. As will be seen in the following section, this is not the case for a broad distribution of pulse velocities.

For an exponential pulse function, the general expression for the cumulants given by \Eqref{eq.cumulants.general} simplifies significantly since the exponential function due to linear damping combines with the pulse function,
\begin{equation} \label{eq.cumulants2}
    \kappa_n(x) = \frac{1}{n\tauw}\,\ave{ a^n\tau \exp{\left( -\frac{nx}{v\taup} \right)} } ,
\end{equation}
where the pulse duration $\tau$ is given by \Eqref{eq.tauk.onesided}. The factor $1/n$ comes from the integration of the $n$'th power of the exponential pulse function. In the case of a degenerate distribution of pulse velocities and amplitudes that are uncorrelated with the pulse durations, the cumulants simplify to
\begin{equation}\label{eq.cumulant.exp}
    \kappa_n(x) = \frac{\taud}{\tauw} \frac{\ave{a^n}}{n} \exp{\left( -\frac{nx}{v\taup}\right)} ,
\end{equation}
where $\taud$ is the pulse duration averaged over the distribution of pulse sizes. It follows that the cumulants and the raw moments decrease exponentially with radius. In particular, the variance is given by
\begin{equation}\label{eq.variance.degenerate}
    \Phirms^2(x) = \frac{\taud}{\tauw} \frac{\langle{a^2}\rangle}{2} \exp{\left( -\frac{2x}{v\taup}\right)} .
\end{equation}
However, the relative fluctuation level $\Phirms/\Phiave$ and the skewness and flatness moments are all constant as function of radius. Additionally assuming exponentially distributed pulse amplitudes as given by \Eqref{eq.pa.exp}, the cumulants are given by
\begin{equation}
    \kappa_n(x) = \frac{\taud}{\tauw} \frac{n!}{n}\left[ \aave \exp{\left( -\frac{x}{v\taup}\right)} \right]^n ,
\end{equation}
which are the cumulants of a Gamma distribution with scale parameter $\ave{a}\exp(-x/v\taup)$ and shape parameter $\taud/\tauw$. The relative fluctuation level and the skewness and flatness moments then become
\begin{align}
    \frac{\Phirms}{\ave{\Phi}} & = \left(\frac{\tauw}{\taud}\right)^{1/2}, \\
    S_\Phi & =  2\left(\frac{\tauw}{\taud}\right)^{1/2}, \\
    F_\Phi & = \frac{6\tauw}{\taud} .
\end{align}
The probability density function for $\Phi$ is the Gamma distribution given by \Eqref{eq.PPhi.Gamma} with the average radial profile given by $\Phiave(x)=\gamma\aave\exp{(-x/v\taup)}$. Thus, in the case of a degenerate distribution of pulse velocities, the shape parameter is fixed but the scale parameter for the distribution decreases exponentially with radius. As will be discussed in \Secref{sec.fpp}, in general, no closed-form expression of the probability distribution function can be obtained in the case of a random distribution of pulse velocities. One notable exception is the case of a discrete uniform distribution of pulse velocities, which will be considered in \Secref{sec.numerical}.

\subsection{Filtered Poisson process}\label{sec.fpp}

A pulse moving with constant velocity $v$ will arrive at a radial position $\xi$ at time $s_{\xi}$ given by
\begin{equation} \label{txik}
    s_{\xi} = s + \frac{\xi}{v} .
\end{equation}
The arrivals $s$ at $x=0$ are assumed to be uniformly distributed on the interval $[-T/2,T/2]$, as described by \Eqref{eq.p_tk.uniform}. In the case of a random distribution of pulse velocities $v$, the arrivals $s_{\xi}$ at position $\xi$ are given by a sum of two random variables and therefore the distribution of these arrivals is given by the convolution
\begin{equation}\label{eq.ptxi}
    P_{s_\xi}(s) = \int_{-\infty}^{\infty} \text{d}r\,P_{\xi/v}(r)P_s(s-r) = \frac{1}{T}\int_{-T/2+s}^{T/2+s} \text{d}r\,P_{\xi/v}(r) ,
\end{equation}
where $P_{\xi/v}$ is the distribution of the radial transit times $r=\xi/v$. It follows that the pulse arrivals at radial position $\xi$ are in general not uniformly distributed. A distribution of pulse velocities leads to end effects that influence the arrival time distribution. This is solely an effect of the radial motion and is independent of the linear damping.

In order to determine the arrival time distribution, consider the case of a velocity distribution $P_v(v)$ that is bounded by a minimum velocity $\vmin$ and a maximum velocity $\vmax$, which results in a maximum transit time $\rmax=\xi/\vmin$ and a minimum transit time $\rmin=\xi/\vmax$, respectively. The probability distribution $P_{\xi/v}(r)$ then vanishes for $r<\rmin$ as well as for $r>\rmax$, and the integral in \Eqref{eq.ptxi} can be rewritten as
\begin{equation} \label{eq:Parrival_xi}
    P_{s_\xi}(s) = \frac{1}{T}\int_{\max(-T/2+s, \rmin)}^{\min(T/2+s, \rmax)} \text{d}r\,P_{\xi/v}(r) .
\end{equation}
Thus, for arrival times $s_\xi$ such that $-T/2+r_\text{max}\leq s\leq T/2+r_\text{min}$ we obtain
\begin{equation}
    P_{s_\xi}(t) = \frac{1}{T}\int_{\rmin}^{\rmax} \text{d}r\,P_{\xi/v}(r) = \frac{1}{T} .
\end{equation}
That is, a broad velocity distribution leading to transit times in the interval $[\rmin,\rmax]$ will result in a distribution of arrival times $s_\xi$ at the radial position $\xi$ that is uniform on the interval $[-T/2+\rmax,T/2+\rmin]$ and therefore constitute a Poisson point process. Note that this assumes $T>\rmax-\rmin$. In the case of a degenerate  distribution of pulse velocities $\rmax=\rmin=\xi/v$ and the pulse arrivals at $\xi$ constitute a Poisson process on the translated interval $[-T/2+\xi/v, T/2+\xi/v]$. For a sufficiently long realization of the process, $T\gg\rmax-\rmin$, end effects can be neglected and the process follows Poisson statistics in the radial domain of interest. Further discussions of end effects and the rate of the process are given in \Appref{app.fpp}.

As discussed above, a pulse $\phi$ will arrive at position $\xi$ at time $s_{\xi}=s+\xi/v$. The superposition of pulses at this position can thus be written as
\begin{equation}\label{eq.Phi_K_xi}
    \Phi_K(\xi,t) = \sum_{k=1}^{K(T)} a_{\xi k} \exp\left( - \frac{t-s_{\xi k}}{\taup} \right) \varphi\left( -\frac{v_k(t-s_{\xi k})}{\ell_k} \right) ,
\end{equation}
where the pulse amplitudes are given by
\begin{equation}\label{eq.axi}
    a_{\xi} = a_0 \exp\left( - \frac{\xi}{v\taup} \right) ,
\end{equation}
with $a_0$ the pulse amplitudes specified at the reference position $\xi=0$. Due to the linear damping and time-independent velocities, the pulse amplitudes decrease exponentially with increasing radial position $\xi$. When the pulse velocities are randomly distributed, the distribution of pulse amplitudes $a_{\xi}$ at $\xi\neq 0$ will be different from the ones specified at the reference position. In particular, the amplitude of slow filaments will decrease substantially with radial position and the process will be dominated by the fast pulses for large $\xi$ since these have shorter radial transit times. As will be discussed below, this correlation between pulse amplitudes and velocities influences the intermittency of the process.

Assuming an exponential pulse function as described by \Eqref{eq:pulse-one-sided}, the exponential amplitude variation can be combined with the pulse function and at the radial position $\xi$ the process can be written as
\begin{equation}\label{eq.phi.fpp}
    \Phi_K(\xi,t) = \sum_{k=1}^{K(T)} a_{\xi k} \varphi\left( - \frac{t-s_{\xi k}}{\tau_k} \right) ,
\end{equation}
where the pulse duration $\tau$ is given by \Eqref{eq.tauk.onesided}. As discussed in \Secref{sec.fpp}, the pulse arrivals $s_{\xi}$ follow a Poisson process when end effects are neglected. The process described by \Eqref{eq.phi.fpp} is therefore a filtered Poisson process generalized to the case of a random distribution of pulse durations. Moreover, this describes how the pulse amplitudes and durations become modified and correlated by a distribution of pulse velocities. Pulses with high (low) velocity will have larger (smaller) amplitudes $a_{\xi}$ and shorter (longer) duration times $\tau$. A distribution of pulse velocities therefore leads to an anti-correlation between amplitudes and durations. The modification of the amplitude distribution and their correlation with pulse durations will be further discussed in \Secref{sec:pamp} for a discrete uniform distribution of pulse velocities.

\section{Discrete uniform velocity distribution}\label{sec.numerical}

The analytical results presented in the previous section show that a distribution of pulse velocities significantly influences both the moments and correlation properties of the stochastic process. Here this will be investigated in detail for the special case of a discrete uniform distribution of pulse velocities, allowing them to take two different values with equal probability,
\begin{equation}\label{eq.Pv.2v}
    P_v(v) = \frac{1}{2}\left[ \delta(v-\vmin) + \delta(v-\vmax) \right] .
\end{equation}
The minimum and maximum velocities are given by $\vmin=(1-\distp)\vave$ and $\vmax=(1+\distp)\vave$, respectively, $\vave=(\vmin+\vmax)/2$ is the average velocity and $\distp$ in the range $0<\distp<1$ is the width parameter of the distribution. The limit $\distp\rightarrow0$ corresponds to the case of a degenerate distribution of pulse velocities. The discrete uniform distribution is presented in \Figref{prof_2v}a) for various values of the width parameter $\distp$. In the following, we present the lowest-order statistical moments and probability distributions, and describe how the statistical properties of the process changes with radial position. These theoretical predictions have recently been confirmed by numerical realizations of the process.\cite{paikina_2022}

Throughout this section, all pulses are assumed to have the same size $\ell$ and we consider for simplicity one-sided exponential pulses with an exponential amplitude distribution at the reference position $x=0$ with mean amplitude $\langle{a_0}\rangle$. As will be seen, closed-form expressions can be derived for all relevant statistical averages and distributions, allowing to analyze and describe all aspects of the process. The process with a random distribution of pulse velocities will be compared to the standard case with a degenerate distribution of velocities, where all pulses have the same velocity $\vave$. The pulse duration is then given by
\begin{equation}
    \tau_* = \frac{\taup\ell}{\vave\taup+\ell} ,
\end{equation}
and the process is Gamma distributed with shape parameter
\begin{equation}
    \gamma_* = \frac{\tau_*}{\tauw} .
\end{equation}
Recall that in this standard case, the average radial profile is exponential, $\Phiave(x)=\gamma_*\langle{a_0}\rangle\exp{(-x/\vave\taup)}$, while higher order normalized moments are radially constant. In particular, the relative fluctuation level is $\Phirms/\Phiave=1/\gamma_*^{1/2}$, the skewness is $S_\Phi=2/\gamma_*^{1/2}$ and the flatness is $F_\Phi=6/\gamma_*$.

\subsection{Radial profiles}

\begin{figure}[tb]
\centering
\includegraphics[width=8cm]{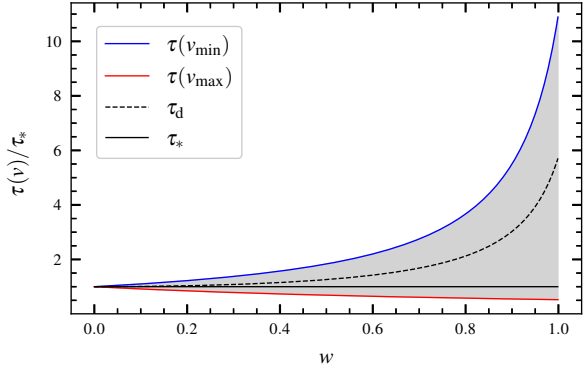}
\caption{Normalized pulse duration $\tau(v)$ for a discrete uniform distribution of pulse velocities as function of the width parameter $\distp$ for $\ave{v}\taup/\ell=10$.}
\label{fig.taud_ave_two}
\end{figure}

The cumulants for the discrete uniform velocity distribution are obtained from \Eqref{eq.cumulants2} by straightforward integration,
\begin{equation}\label{eq.cumulant.2v}
    \kappa_n(x) = \frac{\langle{a_0^n}\rangle}{2n\tauw}\, \left[ \tau(\vmin) \exp{\left( -\frac{nx}{\vmin\taup} \right)} + \tau(\vmax) \exp{\left( -\frac{nx}{\vmax\taup}\right)} \right] ,
\end{equation}
where $a_0$ is the pulse amplitude at the reference position $x=0$ and we have used the notation of a velocity-dependent pulse duration,
\begin{equation}\label{eq.tau(v).2v}
    \tau(v) = \frac{\taup\ell}{v\taup+\ell} .
\end{equation}
The discrete uniform velocity distribution translates into a discrete uniform distribution of pulse durations,
\begin{equation}\label{eq.Ptau.2v}
    P_\tau(\tau) = \frac{1}{2}\left[ \delta(\tau-\tau(\vmin)) + \delta(\tau-\tau(\vmax)) \right] .
\end{equation}
The average pulse duration is given by integration over the discrete distribution,
\begin{equation}\label{eq.taud.2v}
    \taud = \frac{1}{2}\left( \frac{\taup\ell}{\vmin\taup+\ell} + \frac{\taup\ell}{\vmax\taup+\ell} \right) .
\end{equation}
At the reference position $x=0$ the cumulants are given by $\kappa_n(0)=\taud\langle{a_0^n}\rangle/n\tauw$, showing that $\taud/\tauw$ determines the degree of pulse overlap and intermittency at this position. Through $\taud$, the degree of pulse overlap depends on the width of the velocity distribution. The normalized pulse duration $\taud/\tau_*$ is presented in \Figref{fig.taud_ave_two} as a function of the width parameter $w$ for $\vave\taup/\ell=10$. For a fixed average velocity, the average pulse duration increases significantly with the width parameter of the velocity distribution. In fact, in the absence of linear damping, the average pulse duration diverges in the limit $w\rightarrow1$. This is due to the cumulative contribution of nearly stagnant pulses. More generally, the width of the velocity distribution is important for determining the average pulse duration and therefore all cumulants at the reference position.

The width of the velocity distribution also influences the radial variation of the cumulants. In the limit $\distp\rightarrow0$ the velocity distribution tends to a degenerate distribution, giving the familiar exponential profile with scale length $\vave\taup$. For $\distp>0$, the shorter e-folding length of the first term in \Eqref{eq.cumulant.2v} makes this term dominant for negative $x$, while the longer e-folding length of the second term makes this dominant for positive $x$. The cumulants given by \Eqref{eq.cumulant.2v} show that there is a breakpoint $x_\bullet$ between the two exponential functions whose radial location is given by their equal contribution. This depends on the strength of the linear damping and the width of the velocity distribution,
\begin{equation}\label{eq.break_point}
    \frac{x_\bullet}{\ell} = \frac{\vave\taup}{n\ell}\frac{1-\distp^2}{2\distp}\ln{\left( \frac{1+(1+\distp)\vave\taup/\ell}{1+(1-\distp)\vave\taup/\ell} \right)} .
\end{equation}
It is to be noted that the break point is located at positive values of $x$ and decreases with the order of the cumulant. In the limit $\distp\rightarrow1$ the breakpoint approaches the origin. Moreover, the radial location of the breakpoint increases with the normalized linear damping time $\vave\taup/\ell$. Indeed, as discussed previously, in the absence of linear damping these profiles are radially constant and there is no break point.

In general, the statistical properties of the process for negative $x$ are dominated by the slow pulses due to their long radial transit times and therefore excessively large upstream amplitudes, as described by \Eqref{eq.axi}. The process for a large negative position $\xi/\ell$ is effectively a filtered Poisson process given by only the slow pulses,
\begin{equation}\label{eq.Phimin}
    \Phimin(\xi,t) = \sum_{k'=1}^{\Kmin} a_{\xi k'}\varphi\left( - \frac{t-s_{\xi k'}}{\tau(\vmin)} \right) ,
\end{equation}
with the amplitudes given by $a_{\xi}=a_0\exp(-\xi/\vmin\taup)$ and the pulse arrivals $s_{\xi}=s+\xi/\vmin$. This process is Gamma distributed with shape parameter given by $\tau(\vmin)/2\tauw$. Similarly, the statistical properties of the process for large positive $x$ are dominated by the fast pulses since the slow pulses are depleted by the linear damping. Indeed, for a sufficiently large radial position $\xi/\ell$, the process is determined solely by the fast pulses, giving rise to another filtered Poisson process which is given by
\begin{equation}\label{eq.Phimax}
    \Phimax(\xi,t) = \sum_{k''=1}^{\Kmax} a_{\xi k''}\varphi\left( - \frac{t-s_{\xi k''}}{\tau(\vmax)} \right) ,
\end{equation}
with the amplitudes given by $a_{\xi}=a_0\exp(-\xi/\vmax\taup)$ and the pulse arrivals $s_{\xi}=s+\xi/\vmax$. This process is Gamma distributed with shape parameter $\tau(\vmax)/2\tauw$. The dependence of the average pulse duration on the width of the velocity distribution for these two sub-processes is also presented in \Figref{fig.taud_ave_two}, showing how the degree of intermittency varies from far upstream to far downstream of the reference position.

\begin{figure*}[t]
\centering
\includegraphics[width=\textwidth]{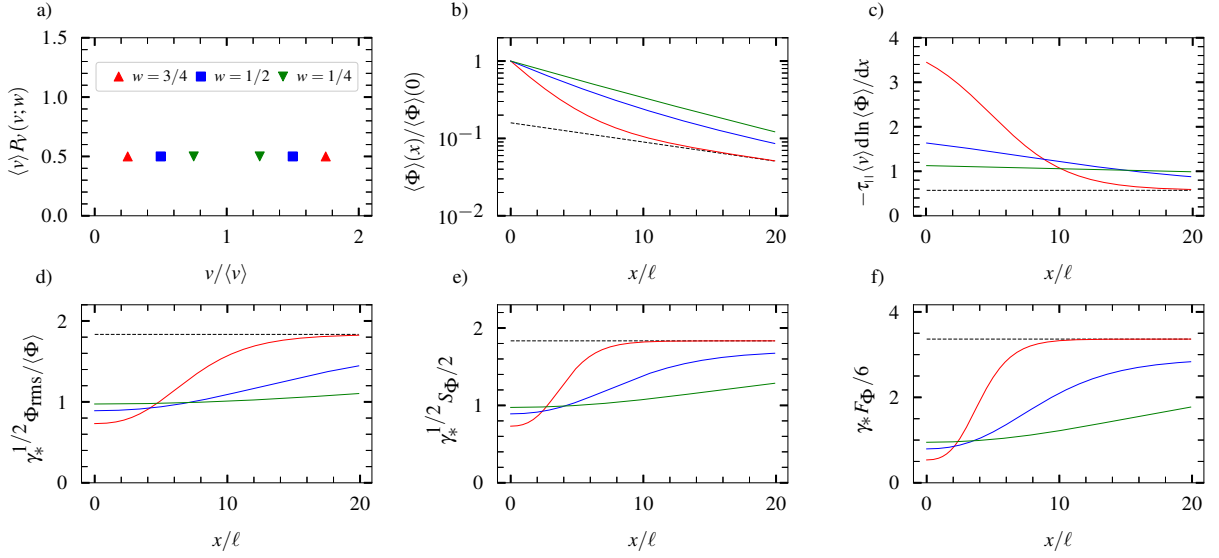}
\caption{Discrete uniform velocity distribution (a) and corresponding radial profiles of the average value (b), inverse profile e-folding length (c), relative fluctuation level (d), skewness (e) and flatness (f) for $\ave{v}\taup/\ell=10$ and various widths $\distp$ of the velocity distribution. All profiles are normalized to their value at the reference position $x=0$ for the base case with a degenerate distribution of pulse velocities for which the intermittency parameter is $\gamma_*=\tau(\vave)/\tauw$. The dashed lines show the asymptotic profiles for large $x/\ell$ corresponding to the process with only the fast pulses.}
\label{prof_2v}
\end{figure*}

From \Eqref{eq.cumulant.2v} it follows that the average radial profile is the sum of two exponential functions,
\begin{equation}\label{eq.phiave.2v}
    \ave{\Phi}(x) = \frac{\langle{a_0}\rangle}{2\tauw} \left[ \tau(\vmin)\exp{\left( - \frac{x}{\vmin\taup} \right)} +  \tau(\vmax)\exp{\left( - \frac{x}{\vmax\taup} \right)} \right] .
\end{equation}
At the reference position this gives $\Phiave(0)=\taud\langle{a_0}\rangle/\tauw$, as expected. The radial profile of the average value $\Phiave$, its normalized inverse e-folding length, the relative fluctuation level and the skewness and flatness moments are presented in \Figref{prof_2v} for $\vave\taup/\ell=10$ and three different values of the width parameter $\distp$. All radial profiles are normalized to their value at the reference position $x=0$ for the standard case of a degenerate distribution of pulse velocities corresponding to $\distp=0$. The breakpoints for the cumulants given by \Eqref{eq.break_point} are indicated by filled circles in \Figref{prof_2v}. For small values of $\distp$, the average profile is nearly exponential and close to that of the reference case, in which all pulses have the same velocity. As expected, the relative fluctuation level, skewness and flatness have weak variation with radial position for small $\distp$. For a wide separation of pulse velocities, the average profile is steep for small $x$ and has a much longer scale length for large $x$, where it is dominated by the fast pulses. Associated with this variation for the average profile is a reduced relative fluctuation level as well as skewness and flatness moments for small $x$, while these quantities increase radially outwards until they saturate at the values associated with the process dominated by the fast pulses, given by the $\Phimax$ process defined above and indicated by the dashed lines in \Figref{prof_2v} for the case $w=3/4$. These profiles demonstrate how a distribution of pulse velocities influences the lowest order moments of the process. In particular, both the relative fluctuation level and the skewness and flatness moments may increase significantly above the levels for a degenerate distribution of pulse velocities.

\subsection{Probability distributions}\label{sec:pamp}

\begin{figure}[tb]
\centering
\includegraphics[width=8cm]{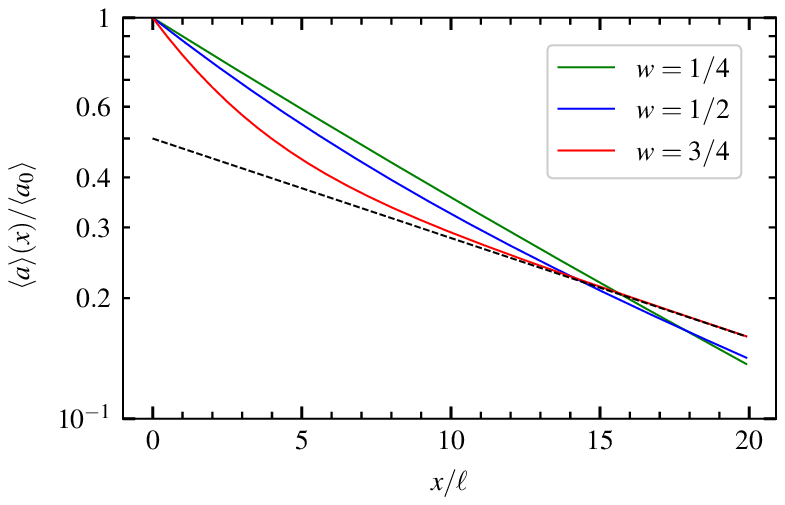}
\caption{Average pulse amplitude as function of radial position for a discrete uniform distribution of pulse velocities with $\vave\taup/\ell=10$ and different values of the width parameter $w$. The dashed line corresponds to the second term in \Eqref{eq.aave.2v}.}
\label{fig.amplitudes_ave_two}
\end{figure}

A non-degenerate velocity distribution will change the amplitude distribution at various radial positions as described by \Eqref{eq.axi}. For the discrete uniform velocity distribution, the radial profile of the average amplitude is given by a sum of two exponential functions,
\begin{equation}\label{eq.aave.2v}
    \aave(x) = \frac{\langle{a_0}\rangle}{2} \left[ \exp{\left( - \frac{x}{\vmin\taup} \right)} + \exp{\left( - \frac{x}{\vmax\taup} \right)} \right] .
\end{equation}
This is presented in \Figref{fig.amplitudes_ave_two} for $\vave\taup/\ell=10$ and three different values of the width parameter $\distp$. For a narrow velocity distribution, the average amplitude decreases nearly exponentially with radial position with scale length $\vave\taup$, similar to the standard case where all pulses have the same velocity. For a wide separation of pulse velocities, the average amplitude decreases sharply with radius for small $x$, while for large $x$ the profile is exponential and dominated by the fast pulses with scale length $\vmax\taup$. This is demonstrated by the dashed line in \Figref{fig.amplitudes_ave_two}, which corresponds to the second term inside the square brackets in \Eqref{eq.aave.2v}.

\begin{figure}[tb]
\centering
\includegraphics[width=8cm]{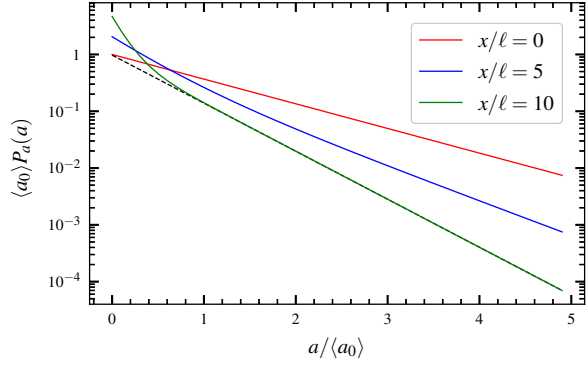}
\caption{Probability density function of the pulse amplitudes for a discrete uniform distribution of pulse velocities with width parameter $\distp=1/2$ at various radial positions in the case $\vave\taup/\ell=10$. The dashed line is the amplitude distribution for the fast pulses.}
\label{fig.amplitudes_pdf_two}
\end{figure}

The probability density function for the pulse amplitudes can be obtained when these are independent of the velocities by using the joint distribution function for the two random variables. The conditional distribution function for the amplitudes at position $x$ given the pulse velocity $v$ is $P_{a\vert v}(a\vert v)$. Since the pulse amplitudes at $x=0$ are exponentially distributed and change with radial position according to \Eqref{eq.axi}, $P_{a\vert v}(a\vert v)$ is an exponential distribution with mean value $\langle{a}\rangle\exp{(-x/v\taup)}$. Since the pulse velocities $\vmin$ and $\vmax$ have equal probability $1/2$, it follows that the probability density function for the pulse amplitudes $a$ at position $x$ with the appropriate normalization is given by
\begin{equation}\label{eq.Pa(x)}
    P_a(a;x) = \frac{1}{2\amin}\exp{\left(-\frac{a}{\amin}\right)} + \frac{1}{2\amax}\exp{\left(-\frac{a}{\amax}\right)} ,
\end{equation}
where we have defined the radial amplitude profile for the fast and slow pulses respectively by
\begin{align}
\amin(x) & = \langle{a_0}\rangle\exp{\left( -\frac{x}{\vmin\taup} \right)} ,
\\
\amax(x) & = \langle{a_0}\rangle\exp{\left( -\frac{x}{\vmax\taup} \right)} .
\end{align}
The amplitude distribution is presented in \Figref{fig.amplitudes_pdf_two} at various radial positions for the parameters $\vave\taup/\ell=10$ and $\distp=1/2$. The distribution is exponential at $x=0$, while for large $x$ the distribution has a clear bi-exponential behaviour with a much higher probability for small amplitudes associated with the slow pulses. The dashed line in \Figref{fig.amplitudes_pdf_two} is the amplitude distribution for the fast pulses, given by the second term in \Eqref{eq.Pa(x)}, showing that the tail distribution is due to the fast pulses.

As discussed above, the process $\Phi_K(x,t)$ for a discrete uniform distribution of pulse velocities can be considered as a sum of the two sub-processes $\Phimin$ and $\Phimax$, each with a degenerate distribution of pulse velocities with values $\vmin$ and $\vmax$, respectively. Accordingly, the probability density function for the summed process is the convolution of the probability distribution of the two sub-processes. Each of these two filtered Poisson processes are Gamma distributed with scale parameter given by the average amplitude $\langle{a_0}\rangle\exp{(-x/v\taup)}$ and shape parameter given by $\tau(v)/2\tauw$ for the two pulse velocities $\vmin$ and $\vmax$, where the pulse duration $\tau(v)$ is defined by \Eqref{eq.tau(v).2v}. The shape and radial variation of the probability density function $P_\Phi$ will depend on the degree of pulse overlap described by $\gamma_*$, the normalized linear damping time $\vave\taup/\ell$, and the width parameter $\distp$ for the velocity distribution. At $x=0$ the distributions of the two sub-processes have the same scale parameter $\langle{a_0}\rangle$, which implies that the probability density function for the summed process is itself a Gamma distribution,
\begin{equation}
    \langle{a_0}\rangle P_{\Phi}(\Phi;x=0) = \frac{1}{\Gamma(\gamma_0)} \left(\frac{\Phi}{\langle{a_0}\rangle}\right)^{\gamma_0-1}\exp{\left(-\frac{\Phi}{\langle{a_0}\rangle}\right)}
\end{equation}
with shape parameter $\gamma_0=\taud/\tauw$. On the other hand, for sufficiently large $x$, the amplitudes of the slow pulses will be depleted due to the linear damping and the process is entirely dominated by the fast pulses, described by \Eqref{eq.Phimax}. In this case, the probability density function for the process will be another Gamma distribution with scale parameter $\langle{a_0}\rangle\exp{(-x/\vmax\taup)}$ and shape parameter $\tau(\vmax)/2\tauw$. For intermediate radial positions, the probability density function is a convolution of two Gamma distributions.

\begin{figure}[ht]
\centering
\includegraphics[width=8cm]{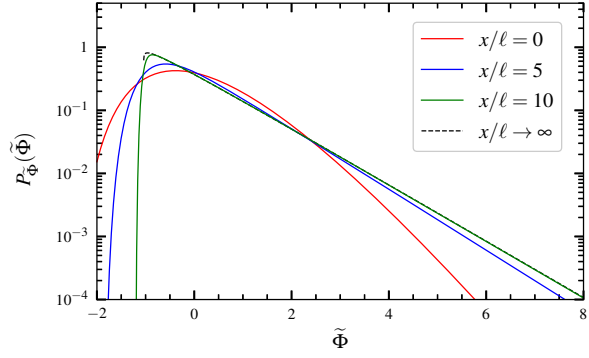}
\caption{Probability density function at various radial positions for intermittency parameter $\gamma_*=20/11$, width parameter $\distp=3/4$ and normalized linear damping time $\ave{v}\taup/\ell=10$. The dashed line is the distribution for the fast pulses.}
\label{fig.Pphi.2v}
\end{figure}

The probability density function $P_{\wt{\Phi}}$ for the normalized variable is presented in \Figref{fig.Pphi.2v} for various radial positions and the parameters $\gamma_*=20/11$, $\vave\taup/\ell=10$ and $\distp=3/4$. At $x=0$ the distribution is unimodal with skewness and flatness moments $S_\Phi=2/\gamma_0^{1/2}\approx1.5$ and $F_\Phi=6/\gamma_0\approx3.3$, respectively, with $\gamma_0=20\left( 1/7 + 1/37 \right) \approx 3.4$. Radially outwards the distribution function becomes strongly skewed and has an exponential tail towards large fluctuation amplitudes. This change in the shape of the probability density function is of course fully consistent with the radial profile of the lowest order statistical moments presented in \Figref{prof_2v}. This demonstrates that a distribution of pulse velocities can lead to significant changes in the probability density function and an increase of the relative fluctuation level and intermittency with radial position. The latter is further emphasized by \Figref{fig.taud_ave_two}, indicating how the intermittency parameter $\gamma_0$ for the process $\Phi_K$ at $x=0$ and for the asymptotic process for large $x$ varies with the width parameter of the velocity distribution.

\begin{figure}[tb]
\centering
\includegraphics[width=8cm]{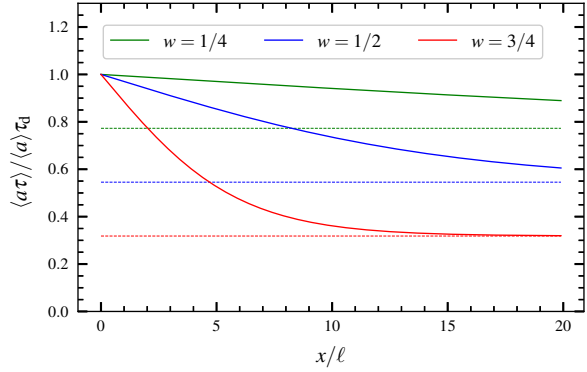}
\caption{Effective pulse duration as a function of the radial position for $\vave\taup/\ell=10$ and different values of the width parameter $\distp$. The dotted lines are the pulse durations $\tau(\vmax)/\taud$ for the fast pulses.}
\label{fig.correlation}
\end{figure}

For the simple case of a discrete uniform distribution of pulse velocities, the process can readily be interpreted in terms of two sub-processes $\Phimin$ and $\Phimax$ corresponding to the two possible velocities as described above. However, as discussed in \Secref{sec.fpp}, a distribution of pulse velocities gives rise to a change in the amplitude distribution and a correlation between pulse amplitudes and durations, which influences the intermittency of the process. Figure~\ref{fig.correlation} shows the radial variation of the linear correlation between pulse amplitudes and durations, $\langle{a\tau}\rangle/\langle{a}\rangle\taud$, for $\vave\taup/\ell=10$ and different values of the width parameter $\distp$. As is clear from \Eqsref{eq.axi} and \eqref{eq.tau(v).2v}, an increasing pulse velocity gives large amplitude and shorter duration, resulting in a significant anti-correlation between these quantities. The quantity presented in \Figref{fig.correlation} can also be interpreted as an effective pulse duration that is weighted with the pulse amplitude. At large radial positions, the process is dominated by the fast pulses, resulting in an effective pulse duration given by $\tau(\vmax)$, which obviously increases the intermittency of the process.

\begin{figure*}[tb]
\centering
\includegraphics[width=\textwidth]{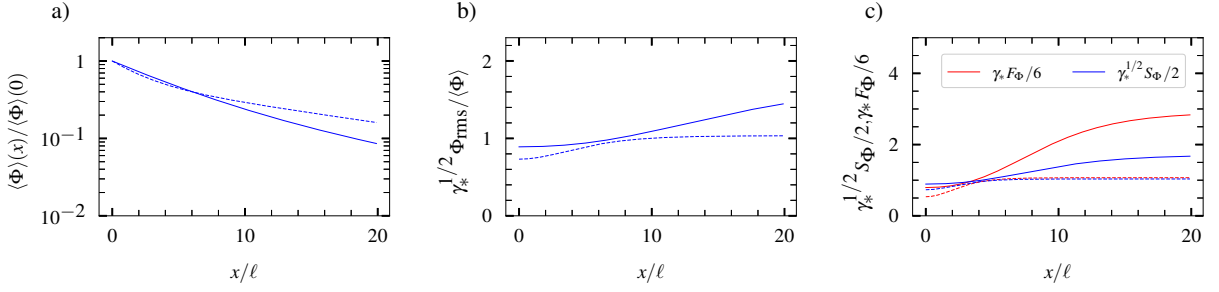}
\caption{Radial profiles of (a) the average value, (b) relative fluctuation level and (c) skewness and flatness moments for $w=3/4$ and $\ave{v}\taup/\ell=10$. All profiles are normalized to their value at the reference position $x=0$ for the standard case with a degenerate distribution of pulse velocities for which the intermittency parameter is $\gamma_*=\tau(\vave)/\tauw$. Full lines are for the original process $\Phi_K$ and broken lines for the modified process $\Psi_K$ where the correlation between pulse amplitudes and durations have been removed.}
\label{fig.prof_correlation}
\end{figure*}

The radial variation of intermittency in the process is due to the change in amplitude distribution with radius, as described by \Eqref{eq.Pa(x)} and shown in \Figref{fig.amplitudes_pdf_two}, as well as the linear correlation between pulse amplitudes and durations. In order to separate these, consider the modified filtered Poisson process
\begin{equation}
    \Psi_K(\xi,t) = \sum_{k=1}^{K(T)} a_{\xi k} \varphi\left( - \frac{t-s_{\xi k}}{\tau_k} \right) ,
\end{equation}
where the arrival times are given by \Eqref{txik}, the pulse durations are distributed according to \Eqref{eq.Ptau.2v} with the mean value given by \Eqref{eq.taud.2v}, and the amplitudes are distributed according to \Eqref{eq.Pa(x)}. This process thus has the same marginal distributions of pulse amplitudes and durations as the original process $\Phi_K$ but with the correlation between amplitudes and durations artificially removed. It is straightforward to calculate the cumulants and the radial profile of the lowest order statistical moments of this modified process. Figure~\ref{fig.prof_correlation} show the radial variation of the lowest order statistical moments for the two processes $\Phi_K$ and $\Psi_K$ for width parameter $w=3/4$ and normalized linear damping time $\vave\taup/\ell=10$. For large $x/\ell$, the original process $\Phi_K$ has significantly higher relative fluctuation level as well as skewness and flatness moments than the process $\Psi_K$ where the correlations have been removed. This is due to the fact that a wide separation of pulse velocities influences both the pulse amplitudes and durations, and the fast pulses with large amplitudes and short durations dominate the process far downstream.

\section{Discussion}\label{sec.discussion}

The statistical properties of a stochastic process given by a super-position of uncorrelated pulses with a random distribution of amplitudes, sizes and velocities have been described. The pulses are assumed to move radially with time-independent velocities and are subject to linear damping, resulting in pulse amplitudes decaying exponentially in time. General results for the cumulants and lowest-order moments of the process have been obtained. When end effects are neglected in realizations of the process and the velocities are time-independent, the rate of pulses remains the same at all radial positions.

In the absence of linear damping, the process is both temporally and spatially homogeneous. Expressions for the cumulants and moments are readily obtained in terms of integrals over the probability distributions. In particular, the cumulants are given by $\langle{a^n\ell I_n/v}\rangle/\tauw$. For an exponential pulse function and exponentially distributed pulse amplitudes independent of the pulse sizes and velocities, the probability density function is a Gamma distribution with scale parameter $\aave$ and shape parameter $\gamma=\taud/\tauw$, where $\taud$ and $\tauw$ are the average pulse duration and waiting times, respectively. Any correlation between pulse amplitudes and duration times will modify this probability density function, as discussed in \Secref{sec.fpp}.

The presence of linear damping drastically modifies the statistical properties of the process, leading to an exponential decay of the pulse amplitudes and therefore radial variation of all statistical averages of the process. In the simple case that all pulses have the same size and velocity, the process results in an exponential radial profile of the cumulants and the lowest order moments with a characteristic scale length given by the product of the pulse velocity and linear damping time, as described by \Eqref{eq.cumulants.general}. A broad distribution of pulse velocities leads to non-exponential profiles and a change in the pulse amplitude statistics and their correlation with pulse durations. Low-velocity pulses will undergo significant amplitude decay during their radial motion, resulting in a strongly peaked downstream amplitude distribution. Moreover, there is an anti-correlation between pulse amplitudes and durations and both these mechanisms give rise to increased intermittency of the process.

The special case of exponential pulses allows to combine the effects of linear damping with the pulse function, providing closed-form expression for many of the statistical averages of the process. When all pulses have the same size and velocity, this is a standard filtered Poisson process at any given radial position with mean pulse amplitude given by $\aave\exp{(-x/v\taup)}$ and pulse duration given by the harmonic mean of the linear damping and radial transit times, as described by \Eqref{eq.tauk.onesided} for one-sided pulses and \Eqref{eq.tauk.twosided} for two-sided pulses. The probability density function is again a Gamma distribution but with the scale parameter decreasing exponentially with radial position.

The implications of a random distribution of pulse velocities are most clearly exemplified by a discrete uniform distribution, allowing the pulse velocities to take two different values with equal probability. Closed-form expressions have here been derived for all relevant fluctuation statistics. This results in a bi-exponential function for the cumulants and radial profiles, which are dominated by the slow pulses for negative $x$ and by the fast pulses for positive $x$. This describes features similar to that typically found in experimental measurements in the scrape-off layer of magnetically confined plasmas, namely a steep near scrape-off layer profile and a flatter far scrape-off layer, as well as a radial increase of the relative fluctuation level and skewness and flatness moments. Accordingly, the probability density function changes shape from a near-normal distribution at the reference position and a strongly skewed Gamma distribution in the far scrape-off layer.

The stochastic model presented here does not describe the formation mechanism for blob structures, only the average radial profile and its fluctuations based on radial motion and linear damping of such structures. In magnetically confined plasmas, the blob structures are generally considered to be formed close to the magnetic separatrix. Once they are formed the blob structures will accelerate and move radially outwards. It should be noted that the case with a wide discrete uniform velocity distribution does indeed describe a steeper near scrape-off layer profile in the region that is dominated by the slow pulses. This resembles the properties of the blob formation region. However, the far scrape-off layer is dominated by the fast pulses, resulting in large-amplitude fluctuations. Additionally, the blob formation process determines the rate of pulses, defined by the average waiting time $\tauw$. This is an input parameter for the stochastic model and its value must be determined by a first-principles based approach.

Characteristic scrape-off layer plasma parameters in medium-sized tokamaks are electron and ion temperatures $T_\text{e}=T_\text{i}=25\,\text{eV}$ and a magnetic connection length from the outboard mid-plane to divertor targets of $10\,\text{m}$. For a typical blob size of $\ell=1\,\text{cm}$ and radial velocity $v=500\,\text{m}/\text{s}$, this gives the normalized linear damping rate $\taup v/\ell=0.1$, a pulse duration of approximately $\taud=20\,\mu\text{s}$ and a profile scale length for a degenerate distribution of pulse velocities of $v\taup=10\,\text{cm}$. This justifies the dominant radial transport parameter used in the analysis presented here and demonstrates that this model is capable to describe strongly flattened far scrape-off layer particle density profiles.

Plasma blobs are three-dimensional, magnetic field-aligned structures which have peak amplitudes at the outboard mid-plane region due to unfavourable curvature in toroidal geometry. The filament temperature and initial extension along the magnetic field may change from one event to another. This motivates the introduction of a random distribution of the linear damping time $\tau_\parallel$. The expressions derived for the cumulants in \Eqsref{eq.cumulants.general} and~\ref{eq.cumulants2} remain valid as far as the statistical average is taken also over the distribution of damping times. The effect of a distribution of $\tau_\parallel$ on the statistical properties of the process is effectively the same as that of the pulse velocity $v$, since it determines the e-folding length $v\tau_\parallel$ of the cumulants. Thus, a stochastic process with a degenerate velocity distribution but a random distribution of damping times will be very similar to the process discussed above. A positive correlation between pulse velocities and linear damping will further enhance the effect of fast pulses since the profile e-folding length $v\taup$ will be larger for those: fast pulses with long damping times will reach the far scrape-off layer without significant amplitude reduction. By contrast, slow pulses with short damping times will have substantial amplitude reduction and contribute little to the far scrape-off layer profile and its fluctuations. Thus, the far scrape-off layer will be highly intermittent and dominated by the fast pulses.

\section{Conclusions and outlook}\label{sec.conclusions}

Broad and flat time-average radial profiles of particle density and temperature in the scrape-off layer of magnetically confined plasmas are generally attributed to the radial motion of blob-like filament structures. Simple theoretical descriptions and transport code modelling describe this by means of effective diffusion and convection velocities, neglecting the intermittent and large-amplitude fluctuations of the plasma parameters in the boundary region.\cite{stangeby_plasma_2000} Recently, some first attempts at describing both the fluctuations and the time-average radial profiles have been presented.\cite{garcia_stochastic_2016,militello_scrape_2016,militello_relation_2016} These are based on a stochastic model describing the fluctuations as a super-position of uncorrelated pulses with a random distribution of amplitudes, sizes and velocities.\cite{garcia_stochastic_2012,kube_convergence_2015,theodorsen_level_2016,theodorsen_level_2018,theodorsen_probability_2018,theodorsen_statistical_2017,garcia_auto-correlation_2017,garcia_stochastic_2016}

In this contribution, we have presented the theoretical foundation for a stochastic modelling of blob-like structures in the scrape-off layer, moving radially with a time-independent velocity but subject to linear damping due to drainage along magnetic field lines. General expressions have been derived for the cumulants and lowest order moments for the process in the case of a general distribution of pulse amplitudes, sizes and velocities, as well as correlations between these. Closed-form expressions for an exponential pulse function provide particularly insightful results, clearly demonstrating how a distribution of pulse parameters influences the statistical properties of the process. Even for the simple case of a discrete uniform pulse velocity distribution, many salient features of experimental measurements are recovered by the model, including distinction between near and far scrape-off layer regions, a broad and flat far scrape-off layer profile, radial increase of the relative fluctuation level, and strongly intermittent far scrape-off layer plasma fluctuations.

The stochastic model promises to be a highly valuable framework for analyzing and describing experimental measurements. In particular, imaging data can be used to estimate blob sizes, velocities and amplitudes and correlations between these. Moreover, the model can also be applied to data from first-principles based turbulence simulations of the boundary region in order to describe and understand the relation between blob statistics and resulting time-averaged profiles. Furthermore, the model can be used to validate model simulations against experimental measurement data. In future work, the model presented here will be extended to describe various continuous velocity distributions as well time-varying pulse velocities. In particular, cases where the pulse velocity is given by the instantaneous pulse amplitude will be considered, with specific scaling relationships predicted by theory for isolated plasma filaments.\cite{krasheninnikov_scrape_2001,dippolito_cross-field_2002,bian_blobs_2003,garcia_mechanism_2005,garcia_radial_2006,myra_collisionality_2006,dippolito_convective_2011,madsen_influence_2011,kube_velocity_2011,higgins_determining_2012,manz_filament_2013,halpern_three-dimensional_2014,easy_three_2014,easy_investigation_2016,held_influence_2016,pecseli_solvable_2016,kube_amplitude_2016,walkden_dynamics_2016,wiesenberger_unified_2017,ross_nature_2019}

\section*{Acknowledgements}

This work was supported by the UiT Aurora Centre Program, UiT The Arctic University of Norway (2020). A.~T.~was supported by Tromsø Research Foundation under grant number 19\_SG\_AT. Discussions with O.~Paikina and M.~Rypdal are gratefully acknowledged.

\appendix

\section{Two-sided exponential pulses}\label{app.two.exp}

\begin{figure}[t]
\centering
\includegraphics[width=8cm]{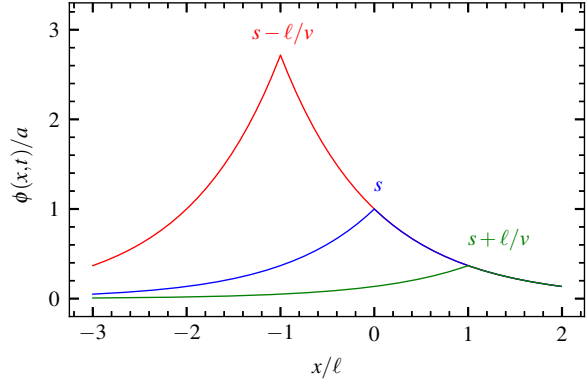}
\caption{Radial variation of a symmetric, $\sigma=1/2$, two-sided exponential pulse at the arrival time $s$ and one radial transit time $\ell/v$ before and after the arrival at $x=0$ for the marginal case $\sigma\ell/v\taup=1$.}
\label{fig.twosided2}
\end{figure}

The case of one-sided exponential pulses can readily be generalized to the case of a continuous, two-sided pulse function,
\begin{equation}\label{eq:pulse-two-sided}
    \varphi(\theta; \sigma) =
        \begin{cases}
            \displaystyle\exp{\left(\frac{\theta}{1-\sigma}\right)}, & \theta \leq 0 , \\
            \displaystyle\exp{\left(-\frac{\theta}{\sigma}\right)},  & \theta > 0 ,
        \end{cases}
\end{equation}
where the spatial pulse asymmetry parameter $\sigma$ is in the range $0<\sigma<1$. For $\sigma=1/2$ the pulse function is symmetric, as shown in \Figref{fig.twosided2}. It is clear that the two-sided exponential pulse contributes to the mean value of the process at any given position $x$ both prior to and after its arrival at this position. The pulse has a steeper leading front than trailing wake for $\sigma<1/2$. In the limit $\sigma\rightarrow0$ this reduces to the simple case of a one-sided exponential pulse function given by \Eqref{eq:pulse-one-sided}. The integral of the $n$'th power of the pulse function, defined by \Eqref{eq.In}, is the same as for one-sided pulses, $I_n=1/n$, independent of the pulse asymmetry parameter $\sigma$.

At any radial position $\xi$ the super-position of pulses can be written as
\begin{equation}\label{eq.PhiK.solution.x0.appendix}
    \Phi_K(\xi,t) = \sum_{k=1}^{K(T)} a_{\xi k}\exp{\left( - \frac{t-s_{\xi k}}{\taup} \right)} \varphi\left( - \frac{v_k(t-s_{\xi k})}{\ell_k} ; \sigma_k \right) ,
\end{equation}
where $s_{\xi}=s+\xi/v$ and $a_\xi$ is given by \Eqref{eq.axi}. For the two-sided exponential pulse function defined by \Eqref{eq:pulse-two-sided} it is straightforward to show that this process can be written as
\begin{equation}\label{eq.PhiK.solution.lambda.appendix}
    \Phi_K(\xi,t) = \sum_{k=1}^{K(T)} a_{\xi k} \varphi\left( - \frac{t-s_{\xi k}}{\tau_k} ; \lambda_k \right) ,
\end{equation}
where the pulse duration is given by the sum of the pulse rise and fall times,
\begin{equation}\label{eq.tauk.twosided}
    \tau = \frac{\taup^2v\ell}{[v\taup+(1-\sigma)\ell](v\taup-\sigma\ell)} ,
\end{equation}
and the temporal asymmetry parameter is the ratio of the pulse rise time and duration,
\begin{equation}\label{eq.twosided.lambda}
    \lambda = \sigma + \sigma(1-\sigma) \frac{\ell}{v\taup} .
\end{equation}
In the limit $\sigma\rightarrow0$ we obtain the case of a one-sided exponential pulse function with vanishing rise time, $\lambda\rightarrow0$, and the pulse duration is the harmonic mean of the linear damping time and the radial transit time given by \Eqref{eq.tauk.onesided}. In the absence of linear damping, the pulse duration is just the radial transit time, $\tau=\ell/v$, and the spatial and temporal asymmetry parameters are the same, $\lambda=\sigma$.

There are some non-trivial criteria for the existence of the mean value and higher-order statistical moments even for a degenerate velocity distribution. From \Eqref{eq.mean.general} it is noted that the pulse function $\varphi$ must decrease sufficiently rapid and at least exponentially for large $\theta$ in order for the integral over $\theta$ to converge. The reason for this possible divergence is that pulses contribute to the mean value and higher order moments at any radial position prior to their arrival at that position when the pulse function is non-negative ahead of the pulse maximum. To illustrate this, consider the two-sided exponential pulse function given by \Eqref{eq:pulse-two-sided}. The average is finite only if $\sigma<v\taup/\ell$, that is, when the weighted radial transit time $\sigma\ell/v$ is shorter than the linear damping time $\taup$. Otherwise, the integral over positive $\theta$ diverges.

The radial variation and evolution of a pulse for the marginal case $\sigma\ell=v\taup$ is presented in \Figref{fig.twosided2} for the arrival time $s_k$ as well as one radial transit time $\ell/v$ before and after the arrival at $x=0$. When $\sigma\ell<v\taup$ the pulse amplitude decay during the radial transit is so weak that the mean value at any radial position is dominated by the leading front from upstream pulses. This leads to a divergence of the mean value of the process as well as all higher-order moments. Clearly, for $0<\sigma<1$ and $0<\sigma\ell/v\taup<1$ the pulse duration given by \Eqref{eq.tauk.twosided} is positive definite. It is to be noted that the requirement $\sigma<v\taup/\ell$ must hold for all pulses in the process, so fast and short length scale pulses set the strongest requirement for the asymmetry parameter $\sigma$. For one-sided exponential pulses, there are no such requirements for the existence of the average.

\section{End effects and Poisson process}\label{app.fpp}

\begin{figure}[t]
\centering
\includegraphics[width=8cm]{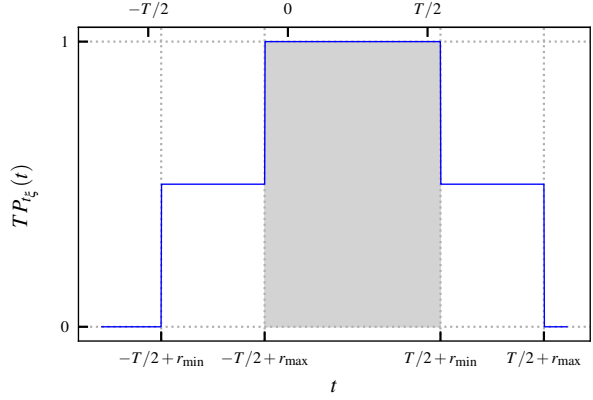}
\caption{Distribution of pulse arrival times $s_{\xi}$ at the radial position $\xi=T\ave{v}/12$ in the case of a discrete uniform distribution of pulse velocities with minimum and maximum radial transit times $\rmin$ and $\rmax$, respectively. The arrival time distribution $P_{s_\xi}$ is uniform in the interval $[-T/2+\rmax, T/2+\rmin]$, indicated by the shaded area in the figure.}
\label{fig.poisson.arrivals}
\end{figure}

The end effects discussed in \Secref{sec.fpp} are clearly illustrated with the example of a discrete uniform distribution of pulse velocities given by \Eqref{eq.Pv.2v}. Assuming $T>r_\text{max}-r_\text{min}$, the pulse arrival time distribution becomes
\begin{equation}\label{eq.Pt(xi)}
    T P_{s_\xi}(t) =  \begin{cases} 
      0 , & s_\xi < - T/2 + \rmin , \\
      \frac{1}{2} , & - T/2 + \rmin < s_\xi < -T/2 + \rmax , \\
      1 , & -T/2 + \rmax < s_\xi < T/2 + \rmin , \\
      \frac{1}{2} , &  T/2 + r_\text{min}  < s_\xi < T/2 + r_\text{max} , \\
      0 & T/2 + \rmax < s_\xi .
   \end{cases}
\end{equation}
This distribution is presented in \Figref{fig.poisson.arrivals} for the case $\xi=T\ave{v}/12$ 
in order to emphasize the presence of end effects. As stated in \Secref{sec.fpp}, the distribution of arrival times is $1/T$ in the range from $-T/2+\rmax$ to $T/2+\rmin$. Neglecting end effects by taking the process duration $T$ to be much larger than $\rmax-\rmin$, the arrival times are uniformly distributed at all radial positions $\xi$ considered. However, the interval of uniform arrivals diminishes as $\vmin$ becomes arbitrarily small, again revealing issues with low pulse velocities. Nevertheless, we conclude that, except for end effects, the pulse arrivals are uniformly distributed at all radial positions and the stochastic process retains its Poisson property with the same rate at all positions. Moreover, it is straightforward to show that the rate of the process is the same as at the reference position $x=0$. It should be noted that based on the results presented here, end effects can easily be accounted for in realizations of the process. These arguments for uniform pulse arrival times do not make any assumptions about the pulse function or distributions of the pulse parameters, only that the pulse velocities are time-independent.

\section{Existence of cumulants}\label{app.cumulants}

Low pulse velocities lead to issues with the existence of cumulants and moments of the process. Upon examination of \Eqsref{eq.cumulants.general} and \eqref{eq.cumulants2} it becomes clear that the expected value of the cumulants may not exist for $x<0$. In particular, consider for simplicity the case of one-sided exponential pulses, a degenerate distribution of sizes, and velocities with a probability distribution $P_v(v)$ which is independent of the pulse amplitudes. With these assumptions, the $n$-th cumulant becomes
\begin{equation}
    \kappa_n(x) = \frac{\taup\langle{a^n}\rangle}{n\tauw} \int_0^\infty \text{d}v\,\frac{P_v(v)}{1+v\taup/\ell} \exp{\left( -\frac{nx}{v\taup} \right)} ,
\end{equation}
where $P_v$ is the marginal distribution of pulse velocities. The integral over pulse velocities may not converge for negative values of $x$. Notice that the fraction $1/(1+v\taup/\ell)$ only takes values between $0$ and $1$ and so does not affect the convergence of the integral. Thus, we examine the convergence of the integral
\begin{equation}
    L = \int_0^\infty \text{d}v\,P_v(v) \exp{ \left(\frac{n \abs{x}}{v\taup} \right) } ,
\end{equation}
for which we use absolute value to emphasize that $x<0$. Making a change of variable defined by $u=1/v$ and using the relation $P_u(u)=(1/u^2)P_v(1/u)$, the integral can be written as
\begin{equation}
    L = \int_0^\infty \text{d}u\,P_u(u) \exp{ \left(\frac{n u \abs{x}}{\taup} \right) } .
\end{equation}
In order for this integral to converge for any radial position $x$ and any cumulant order $n$, the distribution $P_u(u)$ needs to decay faster than exponential for large $u$. Indeed, $P_u(u) \sim \exp(-u)$ for $u\rightarrow\infty$ is not sufficient, since the integral will diverge for sufficiently large $\abs{x}$ or $n$. Therefore, we require at least a stretched exponential behaviour, $P_u(u) \sim \exp(-cu^\zeta)$, for large $u$ for some $\zeta>1$, or equivalently $P_v(v) \sim \exp(-c/v^\zeta)$ for $v \rightarrow 0$ for some constant $c$. For most purposes it is sufficient to impose the simpler condition that finite values of the probability distribution $P_v(v)$ should not reach $v=0$, in other words, there is a minimum velocity $v_{\text{min}}$ such that $P_v(v)=0$ for $v<v_{\text{min}}$.

In summary, care should be taken when using this model to interpret profiles for negative radial positions, $x<0$. The reason for the divergence of cumulants is the dominant contribution of slow pulses. Indeed, in the case of time-independent pulse velocities, we have from \Eqref{eq.amplitude}
\begin{equation}\label{eq.Ak.vs.Xk}
    A(t) = a \exp\left( - \frac{X(t)}{v\taup} \right) ,
\end{equation}
where $X(t)=vt$ is the pulse location at time $t$. With the amplitudes specified as $a$ at the reference position $x=0$, slow pulses will have excessively large amplitudes for negative $x$, resulting in divergence which first arrests higher order cumulants as they have a stronger dependence on the pulse amplitudes. The same condition for convergence of the cumulants applies to two-sided exponential pulses.

\section{Size distribution}\label{app.size}

For completeness, we present here the results for a discrete uniform distribution of pulse sizes in the case where all pulses have the same velocity. Denoting the width parameter for this distribution by $w$, the pulse size probability density function is given by
\begin{equation}
    P_\ell(\ell) = \frac{1}{2}\left[ \delta(\ell-\ellmin) + \delta(\ell-\ellmax) \right] ,
\end{equation}
where $\ellmin=\ellave(1-w)$, $\ellmax=\ellave(1+w)$ and $\ellave=(\ellmin+\ellmax)/2$. According to \Eqref{eq.cumulants2}, a distribution of pulse sizes does not change the radial variation of the cumulants and moments for exponential pulses. When the pulse sizes are independent of the amplitudes, the cumulants for the discrete uniform distribution are given by
\begin{equation}
  \kappa_n(x) = \frac{\taud\langle{a^n}\rangle}{n\tauw}\exp\left( - \frac{nx}{v\taup} \right) ,
\end{equation}
where the average pulse duration is given by
\begin{equation}
    \taud = \frac{1}{2}\left[ \tau(\ellmin) + \tau(\ellmax) \right]
\end{equation}
and the size-dependent pulse duration is $\tau(\ell)=\taup\ell/(v\taup+\ell)$. Thus, a distribution of pulse sizes does not change the radial variation of the moments. In the absence of linear damping, the pulse duration is given by $\ellave/v$, independent of the width parameter for the size distribution.

\section{Non-uniform discrete velocity distribution}\label{app.nonunif}

In this Appendix, we present some results obtained for a generalization of the velocity distribution considered in this manuscript. We consider a non-uniform two-velocity distribution,
\begin{equation}\label{eq.Pv.2v.nonunif}
    P_v(v) = q\delta(v-\vmin) + (1-q)\delta(v-\vmax)
\end{equation}
where $q$ in the range $0\leq q\leq 1$ is the probability that the velocity attains the value $\vmin$. For $q=1/2$ this distribution is identical to \Eqref{eq.Pv.2v}. The cumulants are a straightforward generalization of \Eqref{eq.cumulant.2v},
\begin{equation}
    \kappa_n(x) = \frac{\langle{a_0^n}\rangle}{n\tauw}\, \left[ q \tau(\vmin) \exp{\left( -\frac{nx}{\vmin\taup} \right)} + (1-q)\tau(\vmax) \exp{\left( -\frac{nx}{\vmax\taup}\right)} \right] ,
\end{equation}
where each term is weighted with the probability of the given velocity. 
If the amplitudes of the pulses follow an exponential distribution at $x=0$, the amplitude distribution at radial position $x$ is given by
\begin{equation}\label{eq.Pa(x).nonunif}
    P_a(a;x) = \frac{q}{\amin}\exp{\left(-\frac{a}{\amin}\right)} + \frac{1-q}{\amax}\exp{\left(-\frac{a}{\amax}\right)} ,
\end{equation}
where as before
\begin{align}
    \amin(x) & = \langle{a_0}\rangle\exp{\left( -\frac{x}{\vmin\taup} \right)}
    \\
    \amax(x) & =\langle{a_0}\rangle\exp{\left( -\frac{x}{\vmax\taup} \right)} .
\end{align}
This is known as a bi-exponential distribution with coefficients $q$, $\amin$ and $\amax$. The average amplitude decreases exponentially with radius and is given by $\ave{a}(x) = q\amin(x)+(1-q)\amax(x)$.

\bibliographystyle{custom}
\bibliography{SOL}

\end{document}